\documentstyle[11pt,epsfig,rotating]{article}
\newlength{\dinwidth}
\newlength{\dinmargin}
\begin{document}
\setlength{\unitlength}{1mm}
\begin{titlepage}
\begin{flushleft}
{\tt DESY 95-219    \hfill    ISSN 0418-9833} \\
{\tt November 1995}\\
\end{flushleft}

\vspace*{4.cm}
\begin{center}
\begin{Large}

{\bf Jets and Energy Flow in \\
     Photon-Proton Collisions at HERA} \\

\vspace{1.cm}
{H1 Collaboration}    \\
\end{Large}
\vspace*{4.cm}
{\bf Abstract:}
\end{center}
\begin{quotation}
\renewcommand{\baselinestretch}{1.0}\large\normalsize

Properties of the hadronic final state in photoproduction events with
large transverse energy are studied at the electron-proton collider HERA.
Distributions of the transverse energy, jets and underlying event energy
are compared to $\overline{p}p$ data and QCD calculations.
The comparisons show that the $\gamma p$ events can be consistently described
by QCD models including -- in addition to the primary hard scattering
process -- interactions between the two beam remnants.
The differential jet cross sections
$d\sigma/dE_T^{jet}$ and $d\sigma/d\eta^{jet}$ are measured.

\renewcommand{\baselinestretch}{1.2}\large\normalsize

\end{quotation}
\end{titlepage}

\begin{Large} \begin{center} H1 Collaboration \end{center} \end{Large}
 \begin{flushleft}
 S.~Aid$^{14}$,                   
 V.~Andreev$^{26}$,               
 B.~Andrieu$^{29}$,               
 R.-D.~Appuhn$^{12}$,             
 M.~Arpagaus$^{37}$,              
 A.~Babaev$^{25}$,                
 J.~B\"ahr$^{36}$,                
 J.~B\'an$^{18}$,                 
 Y.~Ban$^{28}$,                   
 P.~Baranov$^{26}$
% ...

\end{flushleft}
\begin{flushleft} {\it
% ...
$ ^{12}$ DESY, Hamburg, Germany$^a$ \\
% ...
$ ^{14}$ II. Institut f\"ur Experimentalphysik, Universit\"at Hamburg,
                                     Hamburg, Germany$^ a$  \\
% ...
$ ^{18}$ Institute of Experimental Physics, Slovak Academy of
                Sciences, Ko\v{s}ice, Slovak Republic$^ f$\\
% ...
$ ^{25}$ Institute for Theoretical and Experimental Physics,
                                                 Moscow, Russia \\
 $ ^{26}$ Lebedev Physical Institute, Moscow, Russia$^ f$ \\
% ...
$ ^{28}$ LAL, Universit\'{e} de Paris-Sud, IN2P3-CNRS,
                            Orsay, France\\
 $ ^{29}$ LPNHE, Ecole Polytechnique, IN2P3-CNRS,
                             Palaiseau, France \\
% ...
$ ^{36}$ DESY, Institut f\"ur Hochenergiephysik,
                              Zeuthen, Germany$^ a$\\
 $ ^{37}$ Institut f\"ur Teilchenphysik,
          ETH, Z\"urich, Switzerland$^ i$\\
% ...
$ ^a$ Supported by the Bundesministerium f\"ur
                                  Forschung und Technologie, FRG
 under contract numbers 6AC17P, 6AC47P, 6DO57I, 6HH17P, 6HH27I, 6HD17I,
 6HD27I, 6KI17P, 6MP17I, and 6WT87P \\
% ...
$ ^f$ Supported by the Deutsche Forschungsgemeinschaft\\
% ...
$ ^i$ Supported by the Swiss National Science Foundation\\}
\end{flushleft}

%
\newpage

\section{Introduction}

\noindent
The interaction of almost real photons and protons is the
dominant process at the electron-proton collider HERA.
A small fraction of these events have large transverse energy $E_T$
in the hadronic final state, measured with respect to the
electron-proton beam axis, or show the formation of jets.

\begin{figure}[b]
\epsfig{file=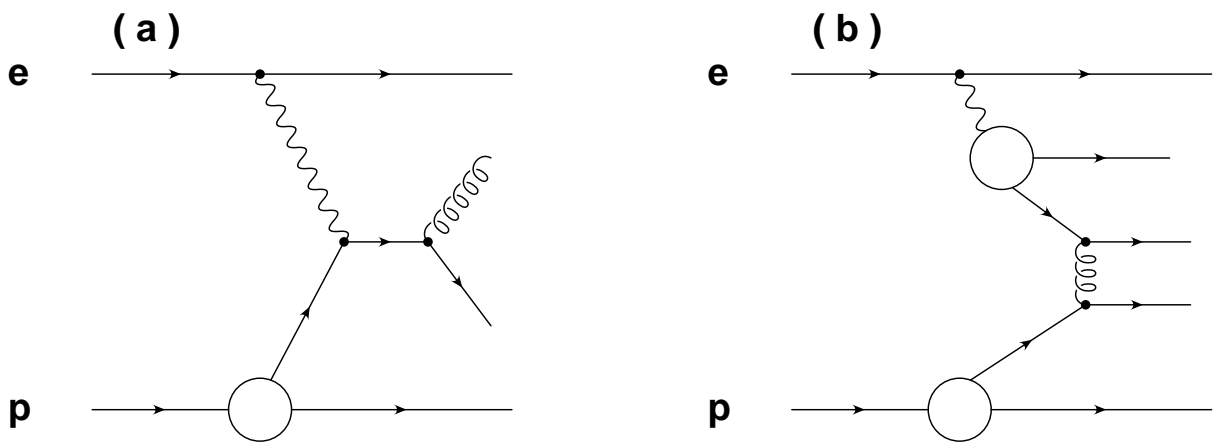,width=16cm}
\caption [$feynman] {\label{feynman}
\small\it Examples of diagrams for direct (a) and resolved photon (b)
processes in electron-proton scattering.
The resolved photon interactions can have --
in addition to the parton scattering process shown --
interactions of the spectator partons of the photon and the proton.}
\end  {figure}

The main characteristics of these events can be described by
perturbative QCD calculations which are usually based on one
hard parton-parton scattering per event.
Two kinds of processes contribute to the high-$E_T$ jet production
in photon-proton collisions (Fig.~\ref{feynman}):
(i)~  direct photon processes, where the photon couples directly to
      a parton in the proton, and
(ii)~ resolved photon processes, in which the scattering occurs
      between a parton from the photon and one in the proton.
Predictions for cross sections are obtained by the convolution
of the matrix elements for the parton scattering with the parton
distributions in the photon and the proton.
Comparisons of measured cross sections with QCD calculations can
provide important information on the
parton scattering processes ~\cite{loqcd}.

Comparisons of HERA data and QCD calculations revealed that the
observed jets are not well described by such
calculations, even if they include phenomenological models
for the QCD radiation effects and the fragmentation phase.
The energy flow adjacent to the jets -- the so called
{\it underlying event energy}, or {\it jet pedestal}  -- was found
to be far above the QCD expectations ~\cite{H1old,zeusjet,gluon,knies}.
This underlying event energy is measured outside of jets
and includes energy resulting from radiation effects of the hard
scattered partons and energy from the fragmentation process of these
partons and the two beam remnants, i.e. the spectator partons.
If interactions between the spectator partons occur in addition
to the hard parton scattering, they also add to the energy level
of the underlying event.
If this underlying event energy is integrated into the measured jet energy,
it  alters drastically the measurement of the jet production rate.
These effects have to be understood before drawing
conclusions on the parton processes
based on the comparison of measured jet cross sections with QCD
calculations.

An excess of the measured underlying event energy above the QCD
calculations has previously been observed in high energy $\overline{p}p$
collisions ~\cite{ellis}.
The data could be described by adding interactions of the
beam remnants to the calculations ~\cite{ua5}.
Since the dynamics of hadronic final state production in photon-proton
interactions is expected to be similar to that in hadron-hadron
collisions, it is interesting to test such models in high energy
photoproduction at HERA ~\cite{schuler,gluon}.

The photon offers a unique probe for the study of the underlying event
energy.
Resolved photoproduction events can in principle involve interactions
between spectator partons of the proton and the photon.
They can be compared to direct photoproduction events,
which by de\-fi\-ni\-tion cannot have these additional interactions.
Such comparisons give information on the underlying event,
as will be shown below.

In this paper detailed studies of photoproduction events with large
transverse energy are presented. Three topics are considered:
\begin{enumerate}
\item The $E_T$ distribution:
      differential cross sections $d\sigma/dE_T$ and
      energy flow in the $\gamma p$ center-of-mass system (cms)
      are analysed, as well as the energy fraction which is
      contained in jets.
\item Properties of the underlying event:
      the energy density outside the jets is determined depending
      on the fraction of the photon energy which is available
      to the hard scattering process.
      Correlations of the transverse energy are measured.
\item Jet shape and rate:
      properties of jets are analysed depending on the jet energy.
      Jet cross sections are presented and the influence of
      the underlying event on the cross sections is studied.
\end  {enumerate}

\section{Detector description and event selection \label{detect}}

A detailed description of the H1 apparatus can be found elsewhere
{}~\cite{H1NIM}.
The following briefly describes the components of the detector relevant
to this analysis, which makes use of the calorimeters,
the luminosity system and the central tracking detector.

The liquid Argon (LAr) calorimeter~\cite{LARC} extends over the polar
angular range
$4^\circ < \theta <  153^\circ$ with full azimuthal coverage, where
$\theta$ is defined with respect to the proton
beam direction ($+z$ axis).
The calorimeter consists of an electromagnetic section with
lead absorbers, corresponding to a depth of between 20 and 30 radiation
lengths, and a hadronic section with steel absorbers.
The total depth of the LAr~calorimeter varies between 4.5 and 8
hadronic interaction lengths. The calorimeter is highly segmented in
both sections  with a total of around 45000 cells.
Test beam measurements of LAr~calorimeter modules have demonstrated
energy resolutions of $\sigma(E)/E\approx 0.12/\sqrt{E}\oplus 0.01$
with $E$ in GeV for electrons ~\cite{H1joerg}
and $\sigma(E)/E\approx 0.5/\sqrt{E}\oplus 0.02$
for charged pions~\cite{H1PI}.
The hadronic energy scale and resolution have been
verified from the ba\-lance of transverse momentum between hadronic
jets and the scattered electron in deep inelastic scattering events
and are known to a precision of $5\%$ and $10\%$ respectively.

The backward electromagnetic calorimeter (BEMC), with a thickness of
22.5 radiation lengths, covers the
region between $151^\circ < \theta < 177^\circ$.
A resolution of $0.10/\sqrt{E}\oplus 0.03$
with $E$ in GeV has been achieved for electrons.
The BEMC energy scale is known to an accuracy of $1.7\%$.

The calorimeter is surrounded by a superconducting solenoid providing a uniform
magnetic field of $1.15$ T parallel to the beam axis in the tracking region.
Charged particle tracks are measured in two concentric jet drift chamber
modules (CJC), covering the polar angular range
$ 15^\circ < \theta < 165^\circ$,
and a forward tracking detector (FTD), covering the
range $ 7^\circ < \theta < 25^\circ$.
The CJC is interleaved with inner and outer layers of multi-wire
proportional chambers, which were used in the trigger to select
events with charged tracks pointing to the interaction region.

The luminosity system consists of two TlCl/TlBr crystal calorimeters having
a resolution of $\sigma(E)/E=0.1/\sqrt{E}$ with $E$ in GeV. The
electron tagger is located at $z=-33$ m and the photon
detector at $z=-103$ m. The electron tagger accepts electrons with an energy
fraction between 0.2 and 0.8 with respect to the beam energy
and  scattering angles below $\theta'\le 5$ mrad $(\theta' = \pi -\theta)$.

The events used in this analysis were taken during the 1993 running period,
in which 26.7 GeV electrons collided with 820 GeV protons in HERA.
The scattered electron was measured in the electron detector of
the luminosity system.
This requirement restricts the negative squared four momentum transfer
of the photon to below $Q^2<0.01$\, GeV$^2$.
Events were selected if they fulfilled the following criteria:
\begin{enumerate}
\item The energy deposited in the electron tagger was in the range
      $8 \le E_{tag} \le 20$ GeV.
      This interval corresponds to scaled photon energies
      $y=E_\gamma/E_e$ between $0.25\le y\le 0.7$.
\item At least one charged particle was measured
      in the central tracker with transverse momentum above
      $0.3$ GeV
      coming from the interaction region.
\item The reconstructed event vertex lay within $\pm 3\sigma$
      of the nominal interaction position along the beam axis.
      The width of the vertex distribution along the beam axis was
      appro\-xi\-ma\-te\-ly Gaussian with $\sigma =10$ cm.
\end{enumerate}
Three different event samples were selected:
\begin{description}
\item{a)}
      ~A sample of events which fulfills the requirements 1,2,3
      is called `minimum bias sample'.
      This sample is only used for comparisons with results
      obtained with the following two samples (see sections 4.2 and 5.1).
\item{b)}
      ~An event sample with large transverse energy in the final state
      (the `high-$E_T$ sample') was defined by \mbox{$E_T\ge 20$\,GeV}
      in the pseudo-rapidity range $-0.8\le\eta\le 3.3$.
      It contained 3254 events and is used in sections 4.1, 4.2, 4.3 and 5.2.
\item{c)}
      ~Events with at least one jet as defined below (the `jet sample').
      This sample contained 3499 events and
      is used in sections 5.1, 6.1 and 6.2.
\end{description}
The pseudo-rapidity in the laboratory system is calculated via
$\eta=-\ln{(\tan{(\theta/2)})}$.
The pseudo-rapidity in the photon-proton cms is calculated via
$\eta^*=\eta-0.5\ln{E_p/E_\gamma}$, where $\eta-\eta^*\approx 2$.
The jet reconstruction was based on purely calorimetric measurements using
a cone algorithm ~\cite{snow} in a grid of cells in the azimuthal angle
$\varphi$ and laboratory pseudo-rapidity $\eta$ which extends from
$-2\le\eta\le 3$.
The cone radius $R=\sqrt{\Delta\eta^2+\Delta\varphi^2}$ was chosen to
be $R=1.0$.
Each jet has transverse energy above $E_T^{jet}\ge 7$ GeV
and lies in the pseudo-rapidity range $-1\le \eta^{jet} \le 2.5$.

The events of samples a) and b) were triggered by a coincidence of an
electron tagger signal and a charged particle measured in the central
proportional chambers.
For all jet analyses, sample c), also one track
from the central jet chamber trigger was required.
Background from non-$ep$ scattering is negligible in all samples.
The integrated luminosity for events with (without) jet requirement
corresponds to $290(117)$\,nb$^{-1}$ determined with an accuracy of $\pm 5\%$.

\section{QCD generators}

\noindent
For comparisons with the data three different event generators were
used.
They are based on tree-level QCD matrix elements.
All calculations were done with the same parton distributions:
for the proton structure the GRV-LO~\cite{pgrv} leading order parton
density parameterizations were used, and
for the photon structure the GRV-LO~\cite{ggrv} leading order
parameterizations.
As will be seen below, the generators can be grouped
into those which include -- in addition to the hard parton scattering --
interactions of the beam remnants and those which do not.
The models for the remnant interactions are based on
parameterizations of results from hadron scattering,
or (semi-~) hard scattering between spectator partons,
or soft and (semi-~) hard parton scattering processes
between spectator partons.

\subsection{PYTHIA}

\noindent
The PYTHIA 5.7 event generator for photon-proton interactions~\cite{pythia}
was used together with a generator for quasi-real photons.
PYTHIA is based on leading order (LO) QCD matrix elements
and includes initial and final state parton radiation calculated in
the leading logarithm approximation.
The strong coupling constant $\alpha_s$ was calculated
in first order QCD using
$\Lambda_{QCD}=200$ MeV with $4$ flavours.
The renormalization and factorization scales were both set to the
transverse momentum $p_t$ of the partons emerging from the hard
interaction.
Since the QCD calculation of a hard parton
scattering process is
divergent for processes with small transverse momenta $p_t$,
the requirement $p_t\ge p_t^{min}=2$\,GeV was made.
For the fragmentation process the LUND fragmentation scheme was used
(JETSET 7.4~\cite{jetset}).
In this PYTHIA calculation the underlying event energy is generated by
initial and final state parton radiation, and by fragmentation
effects.

Within PYTHIA, interactions additional to the primary
parton-parton scattering may be generated, so-called
`multiple interactions' ~\cite{ua5, schuler}.
These are calculated as LO QCD processes between partons from the
photon and proton remnants.
Multiple parton scattering has been studied in proton-antiproton
collisions before ~\cite{ua2mia,afsmia,cdfmia}.
The PYTHIA version extends the concept of the hard perturbative QCD
parton scattering to the low transverse momentum, or semi-hard
interaction region.
As mentioned above, in this low $p_t$ region the LO parton scattering
cross section $\sigma_{parton}$ diverges and becomes larger than the
measured non-diffractive photoproduction cross section $\sigma_{nd}$.
Since in the multiple interaction model each of the two incoming beam
particles may be viewed as a beam of partons,
the problem of too large parton cross sections can be solved by allowing
for several parton scattering processes within one observable $\gamma p$
event.
To prevent a rapid rise of the jet cross section at small transverse
jet energies, a unitarization
scheme has been introduced which results in the damping of the cross
section at small parton momenta.
In this way, the calculated hadronic cross section can stay below the
non-diffractive $\gamma p$ cross section.

In the simplest version of the PYTHIA multiple interaction model
used here,
the transverse momentum cut-off of the hard interactions is
lowered to $p_t^{mia}<p_t^{min}$.
The mean number of \mbox{(semi-)} hard interactions is given by
$<n> = \sigma_{parton}(p_t^{mia})/\sigma_{nd}$, the fluctuations
are calculated from a Poisson distribution.
The number of additional interactions is typically of order $1-2$.
The parton process with the highest transverse momentum in the
partonic final state can be given by any quark or gluon matrix element.
This process includes then initial and final state parton radiation
effects and its partons are connected to the beam remnants by strings.
The additional parton scattering processes in the event are
calculated as perturbative gluon-gluon scattering processes.
The initial state gluon momenta of each subsequent process are
related to the remaining energy of the beam remnants.
The resulting fractional momenta are used to determine the parton
densities of the beam remnants.

The additional interactions contribute significantly to the transverse
energy flow in the event.
This is mainly influenced by the average number of interactions
per event, the cut-off $p_t^{mia}$,
and by the structure functions used for the calculations.
Using the GRV parameterizations for the parton distributions of the
photon and the proton, an adjustment of the transverse momentum
cut-off to $p_t^{mia}=1.2$\, GeV resulted in an adequate description
of the energy flow next to jets as observed in the data
(see section ~\ref{underlying}).
The PYTHIA version with multiple interactions was used to determine
corrections for apparatus inefficiencies.

\subsection{PHOJET}

\noindent
The PHOJET 1.0 event generator was designed to simulate in a
consistent way all components
which contribute to the total photoproduction cross section ~\cite{Ralph}.
It is based on the two-component Dual Parton Model ~\cite{DPM}.
The implementation of the PHOJET generator is similar to the Monte
Carlo event generator DTUJET~\cite{DTUJET} which simulates
multi-particle production in high energy hadron collisions.
This latter generator was originally intended for the description
of soft hadronic interactions and was then extended to hard
scattering processes.
In contrast to PYTHIA,
PHOJET incorporates both, multiple soft and hard parton interactions,
on the basis of a unitarization scheme ~\cite{unitar}.
The soft hadronic processes are described by the soft, `supercritical'
Pomeron ~\cite{supercritical}.
These processes are simulated by a two-string ansatz which allows for
initial transverse momenta of the partons at the ends of the strings.
The hard processes are calculated using the LO QCD matrix elements.
Final state parton radiation effects are simulated using the JETSET 7.4
program ~\cite{jetset}.
Hard initial state parton radiation is not included in this version
of PHOJET.
The lower momentum cut-off for hard parton interactions was set to
$p_t^{min}=3$\, GeV.
Due to the unitarization scheme, small variations of this cut-off
parameter do not have a large influence on the results of this
generator.
The model parameters which describe the soft part of the $\gamma p$
interactions have been tuned using results from proton-antiproton
collisions and low energy photoproduction cross section measurements.
For the fragmentation the LUND string concept is applied
(JETSET 7.4~\cite{jetset}).

\subsection{HERWIG}

The HERWIG 5.8 $ep$ generator is also based on the LO QCD calculations
{}~\cite{herwig}.
This program was designed to have as much input from perturbative QCD
as possible, in order to minimize the free parameters.
HERWIG includes a parton shower model which allows for interference
effects between the initial and final state showers (colour coherence)
{}~\cite{coherence}.
The renormalization and factorization scales were set according
to the transverse momentum of the scattered partons
with a lower cut-off used around $p_t^{min}\ge 2$\,GeV.
The strong coupling constant $\alpha_s$ was calculated to first order using
$\Lambda_{QCD}=180$\,MeV for 5 flavours.
A cluster model is used to simulate the hadronization effects ~\cite{cluster}.

HERWIG allows optionally for additional interactions of the
beam remnants.
These interactions are called {\it soft underlying event} and are
parameterizations of experimental results on `soft' hadron-hadron collisions.
A tuning of the strength and frequency parameters is still in progress.
Recently, also a model for multiple parton interactions has been developed
for HERWIG ~\cite{butterworth}
which has not been used in this paper.

\section{Distributions of transverse energy}

\subsection{Transverse energy cross section}


\noindent
Cross sections as a function
of the total transverse event energy $E_T$ have previously been
measured in $\gamma p$ ~\cite{zeus1} and $\overline{p}p$
scattering at different
cm energies ~\cite{afs,ua1et,cdfet}.
They fall steeply in the low $E_T$ region
(soft hadronic interactions) and tend to flatten towards large transverse
energy.
For sufficiently high $E_T$ the distributions can be equally well
described by either a power law $(E_T)^{-n}$ or an exponential decrease
$\exp{(-\lambda E_T})$.

\begin{figure}[tbp]
\vspace*{-1.5cm}
\epsfig{file=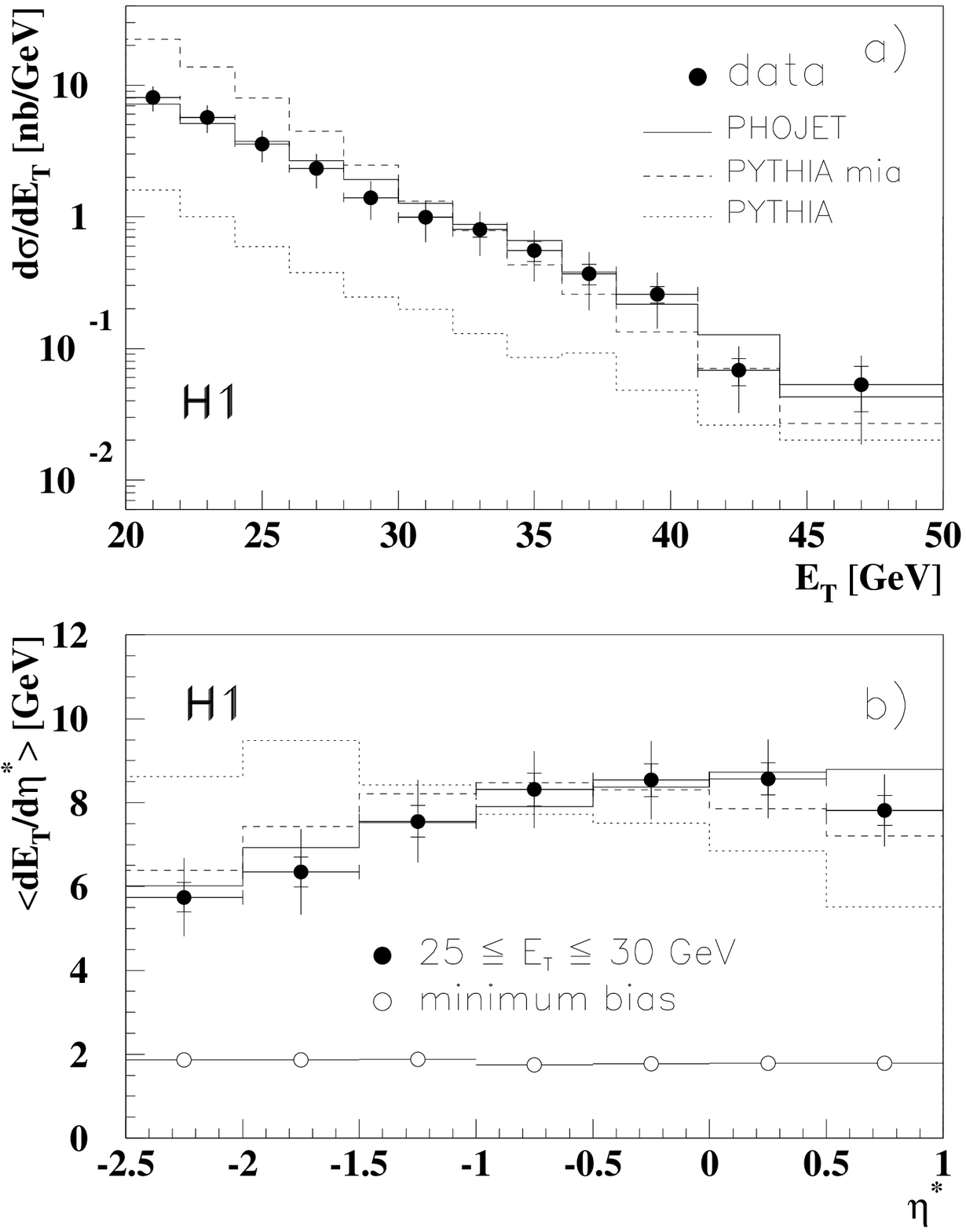,width=16cm}
\caption [$etflow] {\label{etflow}
\small\it
a) The differential transverse energy $ep$ cross section integrated in the
   pseudo-rapidity interval $-2.5\le\eta^*\le 1$.
   Full circles are data, the inner error bars are statistical,
   the outer error bars are the quadratic sum of statistical and
   systematic errors.
   The histograms show the calculations of QCD generators
   with interactions of the beam remnants
   (full=PHOJET, dashed=PYTHIA) and without them (dotted=PYTHIA).

b) ~Corrected transverse energy flow per event versus the
   pseudo-rapidity in the $\gamma p$ cms, where $\eta^*>0$ refers
   to the proton direction.
   Full circles refer to events with total transverse energies between
   $25\mbox{\,GeV}\le E_T(-2.5\le\eta^*\le 1)\le 30$\,GeV.
   Open circles refer to minimum bias data.
   Histograms are as in a).
}
\end  {figure}

In Fig.~\ref{etflow}a the measured differential $ep$ cross section
$d\sigma/dE_T$ is shown.
For this analysis the high-$E_T$ event sample, defined in section
{}~\ref{detect}, was used.
The transverse energy was integrated over the cms pseudo-rapidity range
$-2.5\le\eta^*\le 1$.
This region is fully covered by the electromagnetic and hadronic
sections of the LAr calorimeter.
The measurement corresponds to scaled photon energies between
$0.3\le y\le0.7$ and negative squared momentum transfer
$Q^2\le 0.01$\,GeV$^2$.

Corrections for apparatus inefficiencies applied to
the observed transverse energy were parameterized as
a function of the pseudo-rapidity of the energy deposits.
They decrease with pseudo-rapidity and amount to factors between
$1.2-0.8$.
The corrections depend on the Monte Carlo model
used to determine them. This leads to a contribution
of $10-15\%$ to the systematic error, increasing
with decreasing pseudo-rapidity.
Corrections were applied for the
acceptance of the electron tagger and these
contribute $5\%$ to the systematic error.
This error also includes the uncertainty of the luminosity measurement.
The determination of the trigger efficiency has a $2\%$ error.
The dominant systematic error comes from the uncertainty in the knowledge
of the LAr calorimeter energy scale.
This error increases with $E_T$ and amounts to $15-50\%$.


The differential cross section in Fig.~\ref{etflow}a is compared to
different QCD calculations:
the PYTHIA generator without multiple interactions (dotted histogram)
gives too small values of the cross section.
The HERWIG generator 
gives similar low values (not shown).
The PYTHIA calculation with multiple interactions (dashed histogram)
gives too large cross section values at $E_T=20$\, GeV, but is
compatible with the data in the large $E_T$ region.
The PHOJET generator, which also includes multiple interactions,
provides a good description of the data (full histogram).

The shape of the measured differential $E_T$ cross section can be
described by a power law with $n=5.9$ with a $2\%$ statistical
uncertainty in $n$.
The power $n$ is similar to that of the transverse energy jet cross
section \cite{H1old}.
Therefore the spectrum can be consistently interpreted by
hard parton scattering processes.
The data are in this $E_T$ range equally well described by an
exponential function.
An exponential fit gives the slope
$\lambda=0.21$ at $\sqrt{s_{\gamma p}}=200$\,GeV with a $5\%$ statistical
error in the fit.

\subsection{Transverse energy flow}


\noindent
The peripheral scattering of two hadrons results in a
transverse energy flow
which can be described by a rapidity plateau of width $2\ln{(\sqrt{s}/m)}$,
with final state hadrons of mass $m$ carrying transverse momenta of
around $p_t\approx 300$\,MeV,
$\sqrt{s}$ being the cms energy ~\cite{isrrew}.
In Fig.~\ref{etflow}b the average transverse energy flow in
photoproduction events is shown versus the $\gamma p$ cms
pseudo-rapidity $\eta^*$.
The distributions are corrected for detector effects.
The open circles refer to minimum bias data, defined in section
{}~\ref{detect}.
These data exhibit a plateau within the pseudo-rapidity interval shown,
as expected from peripheral scattering of a hadron-like photon and
a proton.

The full circles refer to the high-$E_T$ event sample, defined in section
{}~\ref{detect},
with corrected total transverse energies, summed
in the pseudo-rapidity range $-2.5\le\eta^*\le 1$,
between $25\mbox{\,GeV}\le E_T\le 30$\,GeV.
The distribution is not compatible with being flat and
increases towards the origin of the $\gamma p$ cms.

Similar distributions have been observed in $\overline{p}p$
scattering with large transverse energy in the final state ~\cite{ua1et}.
Here the $E_T$ distribution versus pseudo-rapidity is centered on the
cms rapidity $\eta^*=0$ and has
(e.g. at $\sqrt{s_{\overline{p}p}}=630$\,GeV)
a half width at half maximum of only one unit of rapidity
for very large $E_T\ge 200$\, GeV which is to be compared with
a possible total plateau width of
$\Delta\eta\sim 13$ (using the proton mass).

The observed increase in the $\gamma p$ cms is, to some extent, unexpected:
since there are relatively more partons at large momentum
fraction $x$ in the photon than in the proton, one would naively
expect this distribution to be peaked somewhere in the photon hemisphere,
i.e. at negative rapidities.
The PYTHIA calculation without multiple interactions (dotted histogram)
confirms this expectation.
A multiple interaction mechanism, however, adds additional energy
around the origin of the $\gamma p$ cms.
The superposition of parton scattering processes by a PYTHIA calculation
with multiple interactions (dashed histogram) results in a description
which is compatible with the data.
Here the maximum is closer to $\eta^*=0$, but still shows a shift
towards the photon hemisphere.
The PHOJET calculation (full histogram), which also includes interactions
of the beam remnants, provides a good description
of the energy flow.

\subsection{Jet rate}


\noindent
In tree level QCD calculations two partons emerge from a hard scattering
process, each fragmenting into a jet.
The measured jet multiplicity depends, however,
on the total transverse energy in the event and
on the jet definition. Here
a cone algorithm is used, with cone size $R=1$
and a transverse energy threshold $E_{T,min}^{jet}$.

\begin{figure} [tbp]
\epsfig{file=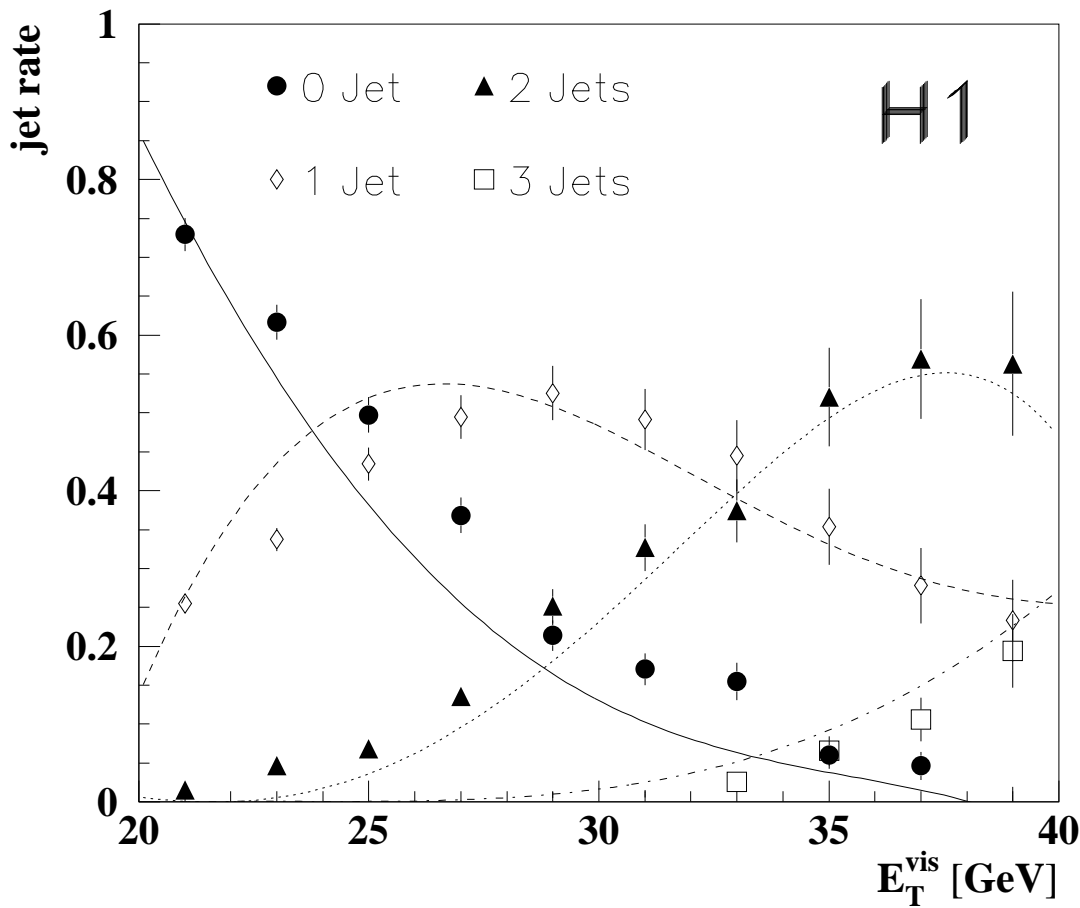,width=16cm}
\caption [$jetrate] {\label{jetrate}
\small\it
Observed relative rate of
0=circles,
1=diamonds,
2=triangles,
3=squares
jet events versus the observed transverse energy
collected in the pseudo-rapidity range $-3\le\eta^*\le 1$.
The observed jet energies $E_T^{jet}$ are above $7$\, GeV summed in a
cone of size $R=1$ for jet axes between $-2.5\le\eta^*\le 0.5$.
Symbols are data with only statistical error bars shown.
The curves are calculations of a QCD generator including interactions
of the beam remnants and a detailed simulation of the detector effects
(PYTHIA).}
\end  {figure}

In Fig.~\ref{jetrate} the relative rates of the observed jet
multiplicities $n_{jet}$ are shown as a function of the uncorrected
transverse event energy $E_T^{vis}$.
These results were obtained using the high-$E_T$ event sample.
The visible jet energy threshold is set to $E_{T,min}^{jet}=7$\,GeV
and the jet axis is required to be in the range $-2.5\le\eta^*\le 0.5$.
The transverse event energy is
the summed transverse energy deposited in the pseudo-rapidity interval
$-3\le\eta^*\le 1$.

At $E_T^{vis}=20$\,GeV, $70\%$ of the events have no jets at all,
given the above definition,
while at $E_T^{vis}=35$\,GeV, half of the events contain two jets.
The two jet configurations are expected from the tree level QCD picture.
At $E_T^{vis}=40$\,GeV more than $10\%$ three jet events are observed, showing
the effect of higher order processes with possible contributions from the
photon remnant.
At this $E_T^{vis}$ there are essentially no events without jets.

The average fraction of the total visible event
transverse energy which is contained in jets
with the chosen parameters was calculated
at $E_T^{vis}=20$\,GeV to be $\sim 10\%$.
At $E_T^{vis}=40$\,GeV about half of the transverse event energy is
contained in jets.

The jet rates are by and large described by calculations of
the PYTHIA generator with multiple interactions, which include here a
detailed simulation of detector effects (Fig.~\ref{jetrate}).
PYTHIA without multiple interactions gives much too large jet rates:
at $E_T^{vis}=30$\, GeV the contribution of events without jets has
already vanished, while the relative two-jet rate is two times the rate
of the data (not shown).
The latter model comparison indicates that the energy depositions
beyond those
resulting from the tree level hard parton scattering process
plus parton
showers are neither correlated with the jets nor jet like, but `soft'
energy depositions spread throughout the event.

\vspace*{1cm}
\noindent
{\bf Summary: Distributions of transverse energy}

\noindent
The photoproduction data with large transverse energy in the
final state show signatures of hard scattering processes:
the slowly decreasing transverse energy spectrum,
the pseudo-rapidity distribution which is not flat,
and the multi-jet production all demonstrate the scattering of
constituents of the photon and proton.
Comparisons of the data with different generators
-- which are based on LO QCD matrix elements plus parton showers --
show that models which include interactions between the beam remnants
provide considerably better descriptions of the data than models
without such a mechanism.

\section{Energy of the underlying event}

\subsection{Transverse energy density outside of jets \label{underlying}}


\noindent

In QCD inspired models various processes contribute to the
transverse energy flow outside jets (e.g. ~\cite{ellis}):
\begin{description}
\item{A.)}
      Transverse energy of the partons that
      participate in the hard scattering
      including 1) initial and 2) final state radiation from these partons,
\item{B.)}
      Transverse energy from the interactions between the spectator
      partons which is
      essentially uncorrelated with the hard scattering process,
\item{C.)}
      Transverse energy of non-interacting spectator partons which is also
      essentially uncorrelated with the hard parton scattering process.
\end  {description}
In all three cases fragmentation effects have to be considered in addition.
Monte Carlo generators which include all these components describe
distributions of the hadronic final state in $\overline{p}p$ ~\cite{ua5}
and $\gamma p$ ~\cite{gluon} scattering considerably
better than generators including only the effects of the hard parton
interaction (item A) and remnant fragmentation (item C).
The clearest experimental evidence for item B would be hard spectator
interactions.
Their observation has been reported in analyses of multi-jet events
in $pp$ and $\overline{p}p$ collisions ~\cite{afsmia, cdfmia}, but
were not confirmed by another $\overline{p}p$ experiment ~\cite{ua2mia}.

At HERA the photon offers a unique probe to study the underlying event:
in $\gamma p$ interactions with {\it resolved} photons the
energy flow outside jets should contain all three components
mentioned above.
{\it Direct} $\gamma p$ interactions, however, do not
involve an initial state parton
from the photon side, hence there is no
initial state QCD radiation (item A.1). There are no
spectator partons interactions (item B) and there is no
photon remnant (item C).
The direct processes should, however, have the same final state
radiation (item A.2) as the resolved photon interactions.
The different contributions can therefore partly be disentangled by
studying the underlying event as a function of, e.g.,
the fractional momentum of the remnant from the photon side
$(1-x_\gamma)$ which is 0
for direct photon interactions and greater than 0 in the case of
resolved photon interactions.


The jet event sample is used (defined in section ~\ref{detect}),
requiring at least two jets.
To avoid a miss-interpretation of the photon remnant as a hard jet
the pseudo-rapidity difference between the two jets had to be below
$\Delta\eta\le 1.2$.
The momentum fraction of the parton from the photon can be
estimated from
\begin{equation}
x_\gamma^{jets} = \frac{E_T^{jet1}e^{-\eta^{jet1}}+E_T^{jet2}e^{-\eta^{jet2}}}
                       {2 E_\gamma}
\label{xg}
\end  {equation}
where $x_\gamma^{jets}$ is reconstructed using the two jets with the
highest transverse energy $E_T^{jet}$ in the event and their
rapidities $\eta^{jet}$.
The energy of the photon $E_\gamma$ is determined from
the energy measured in the electron tagger.
The resolution of $x_\gamma^{jets}$ is between $15-30\%$,
getting worse as $x_\gamma^{jets}$ increases from 0 to 1.

\begin{figure} [tbp]
\epsfig{file=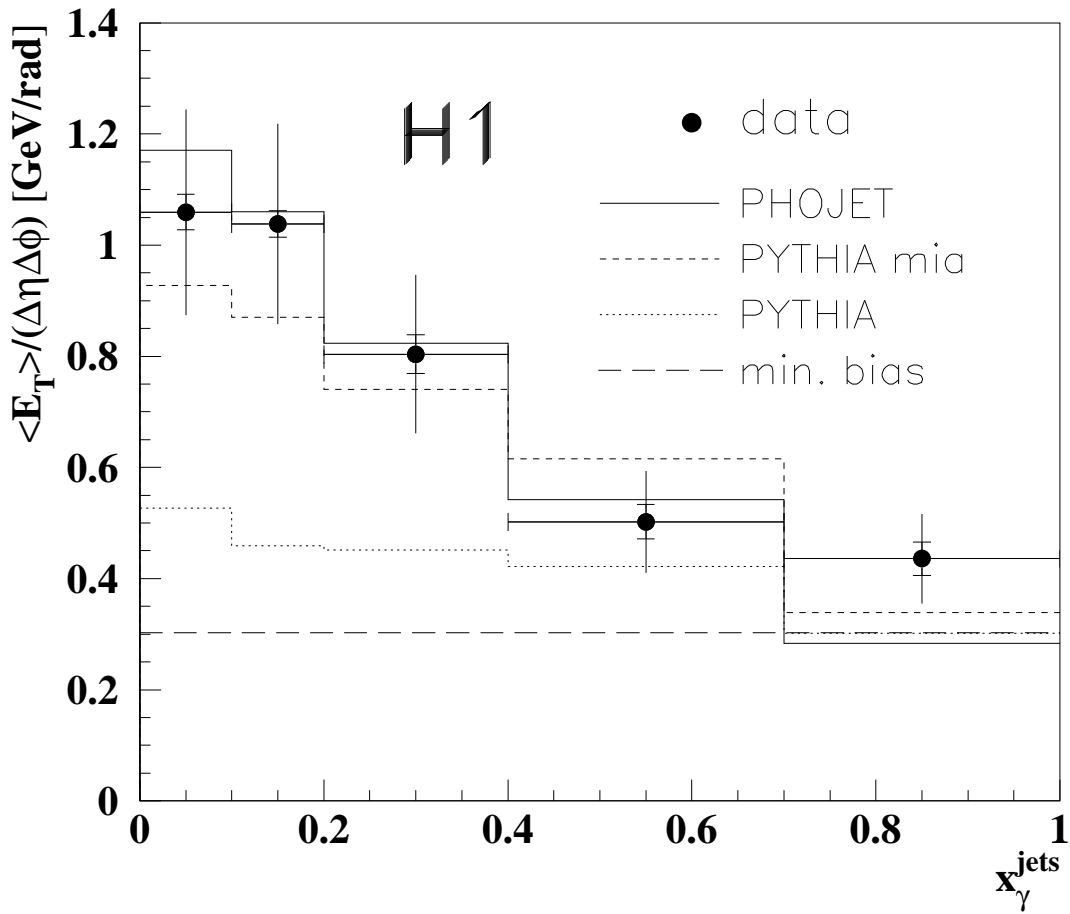,width=16cm}
\caption [$etden] {\label{etden}
\small\it
Corrected transverse energy density in the
central rapidity region
of the $\gamma p$ collision $\vert\eta^*\vert < 1$
per event and unit area in pseudo-rapidity and $\varphi$ space,
outside of the two jets with the highest $E_T^{jet}$.
The variable $x_\gamma^{jets}$ is a measure of the momentum fraction of the
parton from the photon side.
   Circles are data, the inner error bars are statistical, the outer error
   bars result from the quadratic sum of statistical and systematic errors.
   An overall uncertainty in the hadronic energy scale of $5\%$ is
   not shown in the figure.
   The long-dashed line indicates the energy density measured
   in minimum bias events.
   The histograms show the calculations of different QCD generators
   with interactions of the beam remnants
   (full=PHOJET, short-dashed=PYTHIA), and without them (dotted=PYTHIA).
}
\end  {figure}

In this section the measurement of the energy density outside jets
as a function of $x_\gamma^{jets}$ is discussed.
The transverse energy is summed in the
central rapidity region of the $\gamma p$ collision
$-1\le\eta^*\le 1$.
Energy deposited around the two jet axes within
$R=\sqrt{\Delta\varphi^2+\Delta\eta^2}\le 1.3$
is excluded from the energy summation.
The {\it transverse energy density}
$<E_T>/(\Delta\eta\Delta\varphi)$ is then defined as
the energy sum $<E_T>$ per unit area in the
$(\eta,\varphi)$ space averaged over all events in the sample
(Fig.~\ref{etden}).
The distribution was corrected for detector effects.
It rises towards small values of $x_\gamma^{jets}$.

Systematic errors originate from the following sources:
the observed energy was corrected for detector effects, the correction
used depends on the pseudo-rapidity of the energy deposit.
There is a small dependence
on the Monte Carlo model used to obtain the correction
which contributes $10\%$ to the systematic error.
Migration effects in $x_\gamma^{jets}$ were determined with different
generator models and lead to an uncertainty in the energy density
at the level of $14\%$.
These errors are shown in Fig.~\ref{etden}.
In addition, the uncertainty in the knowledge of the hadronic
energy scale of the LAr calorimeter may change the overall
normalization of the energy density by $5\%$.
Contributions from other sources of systematic errors are
negligible.


The long-dashed line in Fig.~\ref{etden} indicates the energy density
found in minimum bias events where $x_\gamma^{jets}$ is not measurable.
The energy density for directly interacting photons ($x_\gamma^{jets}=1$)
is found to be close to the energy density found in minimum bias
events.
This observation is consistent with a recent comparison
of the measured transverse energy density in the
central rapidity region of the $\gamma p$ collision.
Here the same transverse energy density was
measured in minimum bias data and deep inelastic
scattering events (direct photons) ~\cite{andrei}.
These comparisons show that neither the initial state radiation effects
of the parton from the proton, nor the final state parton radiation effects
can be large.
The measured energy density increases towards small $x_\gamma^{jets}$
by $0.6$\,GeV/rad to $3.5$ times the minimum bias energy density.
It is unlikely that such a large effect is caused alone by initial state
radiation of the parton from the photon side.
In the QCD inspired picture described at the beginning of this section,
the natural explanation for the enhanced energy flow is
interactions between the parton spectators (item B):
the probability for these additional interactions increases with
the energy of the photon remnant.

Also shown in Fig.~\ref{etden} are the results of several different Monte
Carlo generators.
The dotted histogram refers to the PYTHIA
calculation without multiple interactions.
This calculation is compatible with the measured energy density
at large $x_\gamma^{jets}$,
but shows only a small increase towards $x_\gamma^{jets}=0$,
as one may expect from increased initial state radiation effects.
The use of another parameterization of the photon structure function,
e.g. LAC1 ~\cite{lac}, does not change this result (not shown).
The HERWIG generator gives similar results (not shown).
The QCD simulations without multiple interactions show that the observed
increase cannot be understood as a kinematical bias.

Adequate descriptions of the data are provided by calculations
which include interactions between the photon
and proton spectator partons, PYTHIA with multiple interactions
(short-dashed histogram) and PHOJET (full histogram).
Using PYTHIA with multiple interactions together with the LAC1
structure function parameterization instead of the GRV-LO set
increases the number of multiple interactions per event due to
the larger gluon density at small fractional momenta $x_\gamma$.
A re-adjustment of the momentum cut-off from
$p_t^{mia}\ge 1.2$\, GeV to $p_t^{mia}\ge 2$\, GeV
results again in an adequate description of the
measured energy density.

For the first time the underlying event energy has been
measured in jet events using direct and resolved photon probes.
The large difference of the energy density in the
central rapidity region of the $\gamma p$ collision
between direct and hadron like photons
and the comparisons with different Monte Carlo
models demonstrate that an additional source of $E_T$
is probably needed beyond that coming from a two parton scattering
process and the according parton radiation effects.

\subsection{Transverse energy correlations}


Energy-energy correlations are sensitive measures of how energy
is distributed over the available phase space.
In this section we examine further the transverse event
energy measured in Fig.~\ref{etflow} and in jet events (Fig.~\ref{etden}).
A study of the rapidity correlations relative to the
central rapidity region of the $\gamma p$ collision,
where the event energy is largest, should
provide information on the underlying event and
is an
important test of the models which describe the average event energy
correctly.

\begin{figure} [tbp]
\epsfig{file=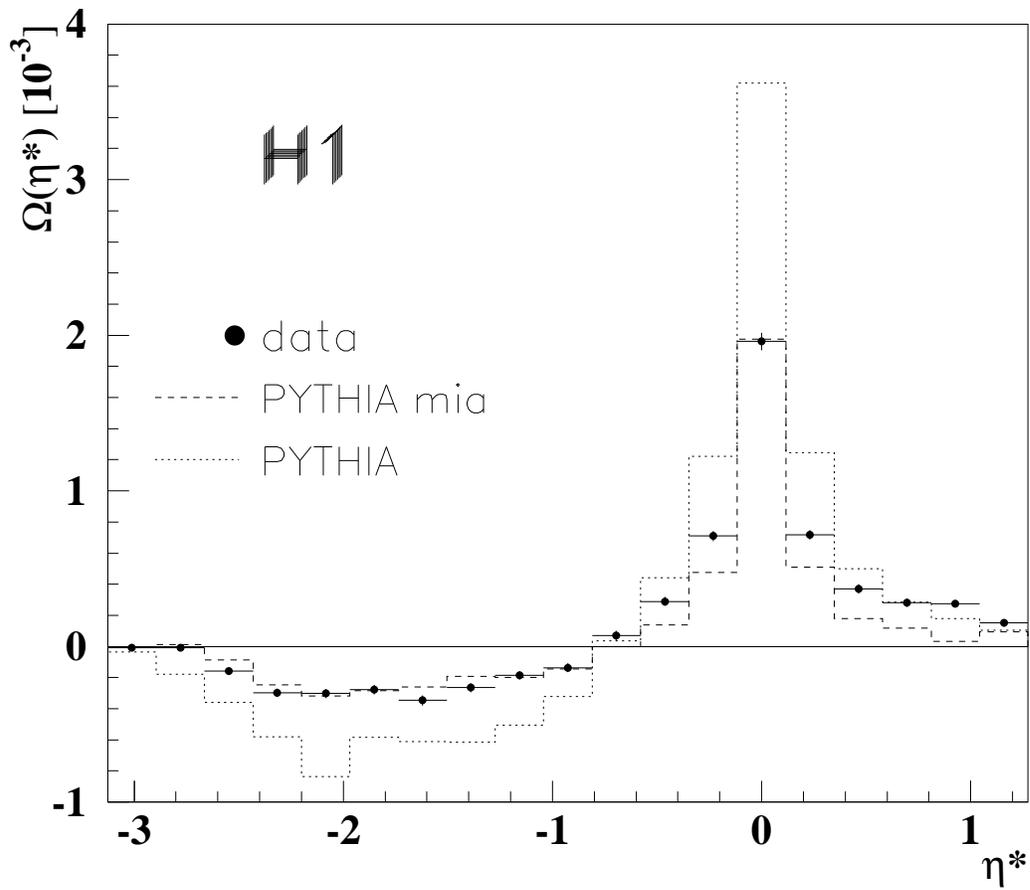,width=16cm}
\caption [$correl] {\label{correl}
\small\it
Observed rapidity correlations with respect to the
central rapidity region of the $\gamma p$ collision
$\eta^*=0$.
Full circles are data, only statistical error bars are shown.
The dashed (dotted) histogram represents a calculation of the QCD
generator PYTHIA with (without) interactions of the beam remnants.
}
\end  {figure}

In Fig.~\ref{correl} rapidity correlations $\Omega$
are shown with respect to the $\gamma p$ cms pseudo-rapidity
$\eta^*=0$ for the high-$E_T$ event sample, defined in section ~\ref{detect}.
Energy deposits within jets are included in this measurement.
The correlation function $\Omega$ was defined as
\begin{equation}
\Omega(\eta^*)=\frac{1}{N_{ev}}\sum_{i=1}^{N_{ev}}
\frac{ \left( <E_{T,\eta^*=0}> - E_{T,\eta^*=0,i} \right)
       \left( <E_{T,\eta^*}> -
         E_{T,\eta^*,i} \right)}
      {(E_T^2)_i} .
\end{equation}
Here $E_T$ is the total transverse energy measured in the range
$-3.1<\eta^*<1.3$, and the other terms refer to transverse energies
measured in the bins of size $\Delta\eta=0.22$.
The average value of the transverse energy in a bin $<E_{T,\eta^*}>$
was determined from all events in the sample.

The data distribution shows short range correlations
around the $\gamma p$ cms reference bin at $\eta^*=0$.
Long range anti-correlations are observed in the photon hemisphere.
These have a minimum around $\eta^*=-1.8$ and disappear around
$\eta^*\approx -3$.


The dotted histogram in Fig.~\ref{correl} shows the calculation
of the PYTHIA generator without multiple interactions,
including a detailed simulation of the detector effects.
The correlations have the correct shape, but the (anti-) correlation
is twice as strong as observed in the data.
This implies that too much energy in the
central rapidity region of the $\gamma p$ collision
is correlated with the other energy deposited in the event.
The PYTHIA version including multiple interactions is shown in
Fig.~\ref{correl} as dashed histogram.
Here event energy is added around $\eta^*=0$ (Fig.~\ref{etflow}b).
This stems from the
interactions of the spectator partons and is not correlated
with the major part of the event energy, which results from the
primary hard scattering process.
Although the model does not give
a perfect description of the data,
it shows that the addition of uncorrelated energy
to the events results not only in the correct average
underlying event energy (Fig.~\ref{etden}), but also gives the
correct correlation strength (Fig.~\ref{correl}).
The same conclusions hold for an event sample where a jet
is explicitly required.

\vspace*{1cm}
\noindent
{\bf Summary: Energy of the underlying event}

\noindent
There are large differences in the
underlying event transverse energy density
in $\gamma p$ collisions dependent on whether the photons involved
are direct or hadron like.
In the
central rapidity region of the $\gamma p$ collision,
processes where
large fractions $x_\gamma\approx 1$ of the photon energy are carried
into the hard scattering process have a similar transverse energy density as
minimum bias events.
Processes with resolved photons $x_\gamma\approx 0$ are found to have
$3.5$ times the transverse energy of minimum bias events.
This increase is much larger than expected from QCD generators
using LO QCD matrix elements plus parton showers, but can be
explained by models including interactions between the beam remnants.
In addition, comparisons of the measured energy-energy correlations
with different models demonstrate that the additional energy of
the beam remnant interactions decreases the correlation strength
to the level observed in the data.

Therefore, the underlying event in photoproduction events can be
consistently interpreted as the superposition of a hard scattering
process plus interactions between the beam spectators.
The characteristic kinematic quantity is the momentum fraction of
the parton from the photon side.
This determines the energy in the hard interaction
and the energy left for interactions of the two beam remnants.

\section{Properties of jets and jet cross sections}

\subsection{Jet shape}


The distribution of transverse energy around a jet axis shows
an approximately Gaussian shape.
In perturbative QCD the width of these jets is expected to decrease with
increasing jet energy.

\begin{figure} [tbp]
\vspace*{-2.0cm}
\epsfig{file=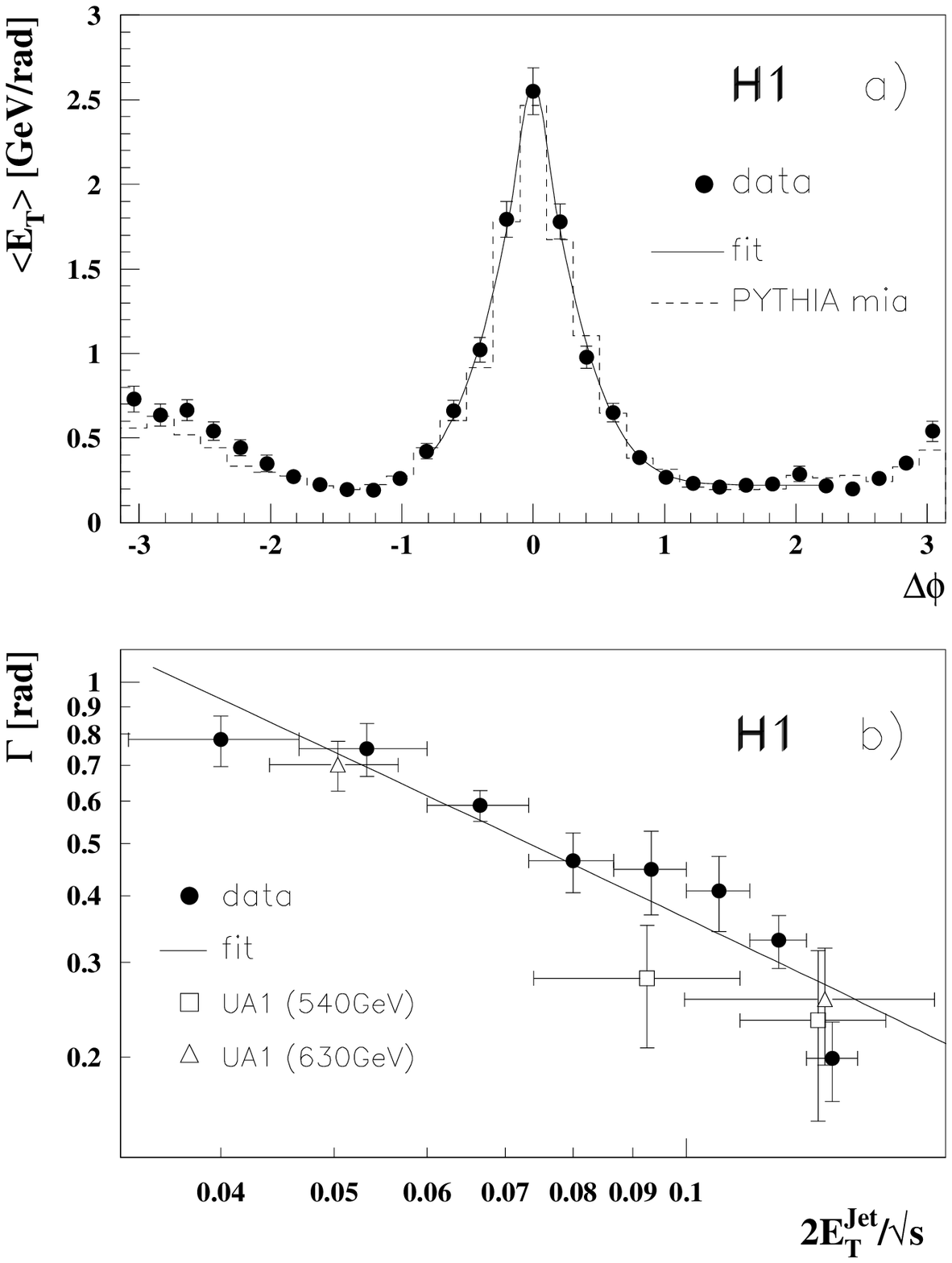,width=21cm}
\caption [$jetprof] {\label{jetprof}
\small\it
a) Observed transverse energy around the jet axis is shown
   versus the azimuthal angle $\Delta\varphi$ for jets
   in the photon hemisphere $-2\le\eta^*_{jet}\le -1$
   and $9\le E_T^{jet}\le 11$ GeV.
   The full circles represent the data with their statistical error bars.
   The axis of the second jet is between $-\pi\le\Delta\varphi\le 0$.
   The histogram is a calculation of the QCD event generator PYTHIA with
   multiple interactions, and the curve is a fit using eq.(~\ref{proffit}).

b) The fitted full width $\Gamma$ at half maximum above the underlying
   event of a jet is shown depending on the scaled jet energy
   $2 E_T^{jet}/\sqrt{s_{ep}}$,
   where the $ep$ center-of-mass energy is used following ~\cite{kramer}.
   Full circles are H1 $\gamma p$ data,
   the error bars represent the quadratic sum of the
   statistical and systematic errors.
   Open symbols are derived from fits to jet profiles of
   $\overline{p}p$ data using the cms energies $\sqrt{s_{\overline{p}p}}$
   ~\cite{ua1ped,ua1prof}.
   The full line represents a $1/E_t^{jet}$ fit to the $\gamma p$ data.
}
\end  {figure}

In Fig.~\ref{jetprof}a the uncorrected transverse energy flow
in the azimuthal direction $\Delta\varphi$
around the jet with the highest $E_T^{jet}$ in events
with at least two jets is shown.
The jets were found in the photon hemisphere between
$-2\le\eta^*_{jet}\le -1$ where detector corrections are small.
The transverse jet energies were between
$9\le E_T^{jet}\le 11$ GeV summed in a cone of size $R=1$.
The energy flow was integrated in a slice of
$\vert\eta^*-\eta^*_{jet}\vert\le 1$ around the jet axis.
The two jets usually deviate from a back-to-back
configuration in the azimuthal angle.
The azimuthal angle between the two jets was defined such that
$-\pi\le\Delta\varphi\le 0$.
Since the jets have a total width of approximately two units
in the azimuthal direction (Fig.~\ref{jetprof}a),
the region $1\le\Delta\varphi\le 2$ is essentially
free of the second jet and shows the underlying event energy.

The full curve in Fig.~\ref{jetprof}a represents an empirical
3 parameter fit of the jet profile using a function closely
related to a Gaussian:
\begin{eqnarray}
f(\Delta\varphi) &=& A
\exp{( - (\sqrt{\vert\Delta\varphi}\vert + b ) ^4 + b^4)} + P
\label{proffit}
\end  {eqnarray}
Parameter $A$ describes the amplitude of the jet at $\Delta\varphi=0$
which is mainly constrained by the energy measured in the central bin.
$P$ reflects the underlying event energy (jet pedestal).
The full width at half maximum above the pedestal is then
$ \Gamma = 2 (( \ln{2}+b^4 ) ^{1/4} - b )^2 $.
Regions potentially affected by the second jet,
$\Delta\varphi<-1$ and $\Delta\varphi>2.2$, are excluded from the fit.

In Fig.~\ref{jetprof}b the width $\Gamma$ of jets from $\gamma p$
interactions (full circles) is shown as a function of the scaled
transverse jet energy  $2 E_T^{jet}/\sqrt{s_{ep}}$.
Here the $ep$ cms energy is used, following a QCD analysis
of jet shapes ~\cite{kramer}.
The decrease of $\Gamma$ for jet energies between
$5\le E_T^{jet}\le 20$\,GeV is clearly seen.
This effect is well described by the QCD generators,
although it is not inconceivable that other models could explain
this effect.
The decrease can be described by a $1/E_T^{jet}$ dependence.
Such a dependence is expected from a QCD analysis presented
in ~\cite{kramer}.
Within errors, no dependence of $P$ on $E_T^{jet}$ was found in this
restricted pseudo-rapidity range.
The amplitude $A$ varies as a function of $E_T^{jet}$ and $\Gamma$.
Systematic errors in Fig.~\ref{jetprof}b ($\le 10\%$)
are dominated by the choice
of the $\Delta\varphi$ interval used for the fit.
Fits to the jet profiles in the rapidity projection give compatible
results within error bars.
Since the jet pedestal increases as a function of the rapidity
in the $\eta$ region considered here, more fit parameters are needed
to describe the data.
This leads to larger error bars on the fitted jet width than in the case
of the fits to the $\varphi$-projections of the jet profiles shown in
Fig.~\ref{jetprof}b.


In the same Fig.~\ref{jetprof}b, the photoproduction data are compared
to results from fits to jet profiles of $\overline{p}p$ data
at different cms energies (open symbols) ~\cite{ua1ped,ua1prof}.
Here pseudo-rapidity projections of the jet profiles were used,
allowing $P$ in (\ref{proffit}) to depend linearly on the rapidity.
The same jet width as in $\gamma p$ data was measured
in $\overline{p}p$ data at different jet energies.
Within the error bars the jet widths found in $\gamma p$ and $\overline{p}p$
are compatible with having the same dependence on the ratio of the jet
energy and the cms energy.

\subsection{Jet cross sections}


\noindent
Differential inclusive jet cross sections have previously been
measured in
photoproduction events at HERA ~\cite{H1old,zeusjet}.
The jet cross section $d\sigma/dE_T^{jet}$ as a function of
transverse energy
falls steeply and can be described by a power law
$(E_T^{jet})^{-n}$.
Such a falling distribution is expected from QCD calculations.
Here the matrix elements of different parton scattering processes
(quark-gluon, gluon-gluon etc.) are summed according to the
quark and gluon distributions in the photon and the proton.
The calculations are only modestly sensitive to the parton distributions,
and reflect essentially features of the matrix elements.
The measured cross section can be described by these calculations
and can, therefore, be interpreted as the result of a parton
scattering processes.
The observed jets are mainly located in the photon hemisphere and
cover there several units in pseudo-rapidity.
Jet cross sections $d\sigma/d\eta^{jet}$ as a function of rapidity
are sensitive to the parton distributions in the photon.
These parton distributions can be extracted by comparison of the data
with QCD calculations, using an iterative procedure.
However, the underlying event energy, explained in section
{}~\ref{underlying}, has a large influence on the measured jet cross
sections, as will be shown below.
Before conclusions may be drawn
on the photon structure, the underlying
event has to be described by the QCD calculation
(see Fig.~\ref{etden}),
or the jet data need to be corrected for underlying event effects
(e.g. ~\cite{gluon}).

\begin{figure} [tbp]
\vspace*{-1.5cm}
\epsfig{file=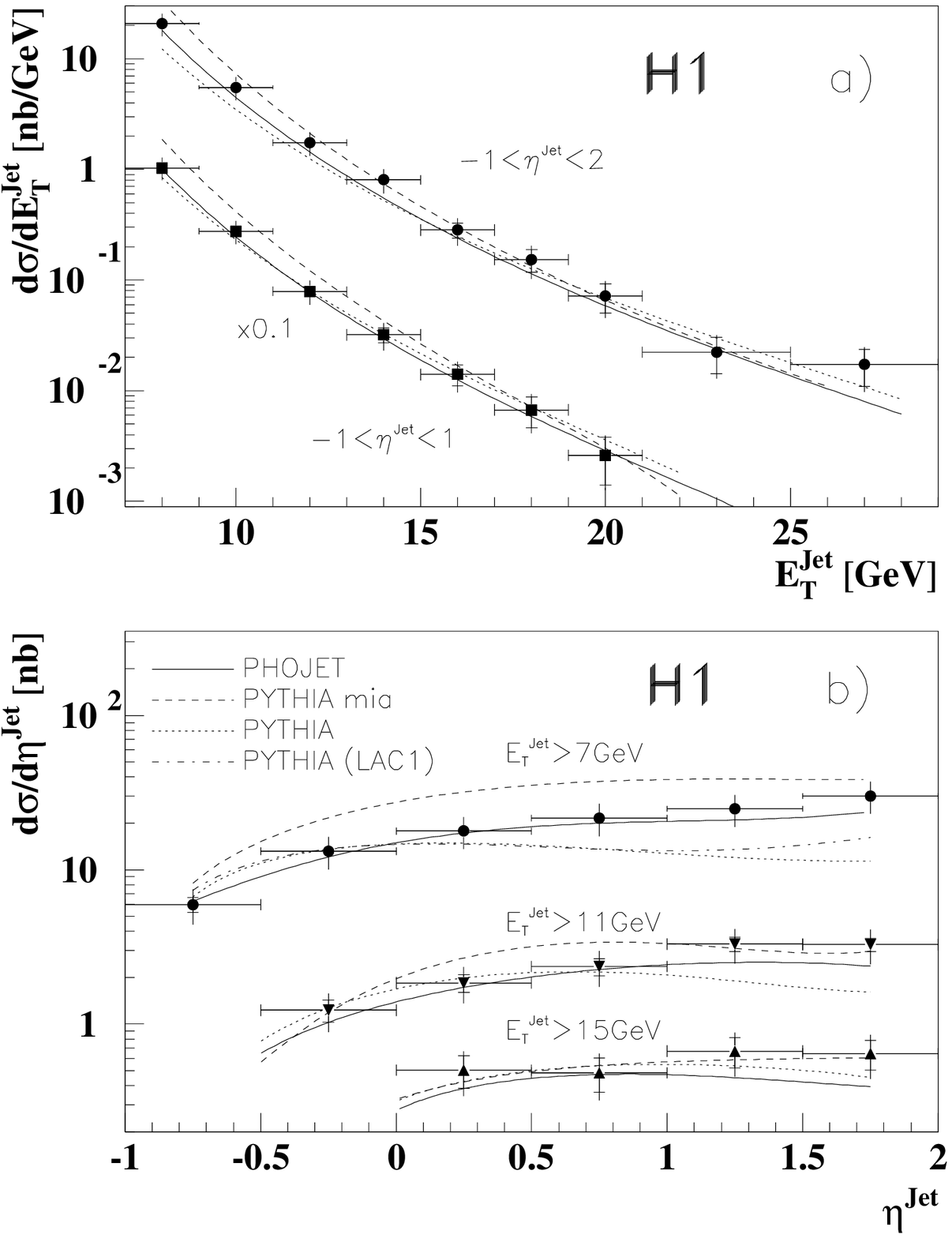,width=16cm}
\vspace*{-1.0cm}
\caption [$sigmajet] {\label{sigmajet}
\small\it
a) Inclusive differential jet $ep$ cross sections for jets with transverse
   energy above $E_T^{jet}  >~ 7$\,GeV summed in a cone of $R=1$.
   Circles are H1 data integrated in the pseudo-rapidity range
   $-1 < \eta^{jet} <  2$,
   squares refer to the region $-1 < \eta^{jet} <  1$ (lowered by a
   factor 10).
   The inner error bars represent the statistical, the outer errors are
   the quadratic sum of statistical and systematic errors which depend
   on $E_T^{jet}$ or $\eta^{jet}$.
   The uncertainty in the overall normalization amounts to $26\%$
   and is not shown in the figure.
   The curves show the calculations of different QCD generators
   with interactions of the beam remnants
   (full=PHOJET, dashed=PYTHIA), and without them (dotted=PYTHIA).

b) Differential jet cross section versus pseudo-rapidity for different
   thresholds in the jet transverse energy:
   circles are H1 data with         $E_T^{jet} >~  7$\,GeV,
   down-pointing triangles with     $E_T^{jet} >~ 11$\,GeV,
   and up-pointing triangles with   $E_T^{jet} >~ 15$\,GeV.
   The dash-dotted curve is a calculation using PYTHIA without
   interactions of the beam remnants, but another parameterization
   of the photon structure function (see text).
   Other curves are described in a).
}
\end  {figure}

In Fig.~\ref{sigmajet} the inclusive differential jet $ep$ cross sections
$d\sigma/dE_T^{jet}$ and $d\sigma/d\eta^{jet}$ are shown.
The measurements were made using the jet event sample.
No corrections for the underlying event energy were applied.
Here the pseudo-rapidity $\eta^{jet}$ in the laboratory system
($\eta-\eta^*=0.5\ln{(E_p/E_\gamma)}\approx 2$) is chosen
for compatibility with previous measurements.
The cross sections are for a
scaled photon energy range $0.25 < y < 0.7$
and negative squared momentum transfer $Q^2 < 0.01$\,GeV$^2$.
The jet cross section as a function of the corrected transverse
jet energy in Fig.~\ref{sigmajet}a was
measured in two pseudo-rapidity intervals $-1 < \eta^{jet} <  1$ and
$-1 < \eta^{jet} <  2$ (see also tables ~\ref{incjetstab1},~\ref{incjetstab2}).
The jet cross section as a function of pseudo-rapidity
in Fig.~\ref{sigmajet}b was measured for
three different thresholds in the jet energy,
$E_T^{jet} > 7, 11, 15$\,GeV
(see also tables ~\ref{incjetstab3},~\ref{incjetstab4},~\ref{incjetstab5}).
The previous measurement of $d\sigma/d\eta^{jet}$ for jet energies above
$E_T^{jet} > 7$\, GeV ~\cite{H1old} suffered from a defect in the
acceptance correction for the small angle electron detector
and is superceded by this new measurement.

Corrections to the energy response of the calorimeter were
made as functions of the transverse energy and pseudo-rapidity
of the jets.
Using generator events the corrections $\epsilon$ result from the
comparison of jet rates resulting from generated hadrons and jets
found after a detailed simulation of detector effects.
The correction function $\epsilon(E_T^{jet})$ decreases from
$1.1$ to $0.7$
as $E_T^{jet}$ increases between $7 <  E_T^{jet} <  30$\,GeV.
The correction $\epsilon(\eta^{jet})$ decreases from $1.7$ to $0.7$
as $\eta$ increases
from $-1$ to $2$.
The systematic uncertainty in the determination of the jet rate
corrections is $10\%$,
including a small dependence on the generator model used.
The hadronic energy resolution is known to the $10\%$ level.
This was found to give a $10\%$ contribution to the systematic error.
The hadronic energy scale of different calorimeter segments
is known to be the same to $3\%$.
The latter results in an $15\%$ contribution to the jet cross section error.
The efficiency of the drift chamber trigger (see section ~\ref{detect})
varies as
function of $E_T^{jet}$ and $\eta^{jet}$
between $90\%$ and $95\%$, with an uncertainty of $3\%$.
In Fig.~\ref{sigmajet} the total error bars include
the quadratic sum of the statistical and systematic errors
mentioned so far.

Sources of systematic errors affecting the cross section
normalization are as follows:
1) ~The dominant systematic error of $25\%$ results from the
    uncertainty of $\pm 5\%$ in the LAr energy scale.
2) ~The error in the luminosity measurement, which includes the
    uncertainty in the electron trigger efficiency, is $5\%$.
These errors have been added in quadrature and give a $26\%$ overall
systematic error in addition to the errors shown in Fig.~\ref{sigmajet}.


The fit to a power law $(E_T^{jet})^{-n}$ of the jet
cross section as a function of transverse energy
in Fig.~\ref{sigmajet}a gives $n=6.1\pm 0.5$
for the pseudo-rapidity interval $-1 < \eta^{jet} <  2$,
where the error reflect the statistical and systematic
uncertainties of the fit.
The measured power $n$ is compatible with previous results
on jet production in $\gamma p$ interactions ~\cite{H1old,zeusjet}.
This result compares also well with jet cross sections measured in
$\overline{p}p$ collisions at the same cms energy
$\sqrt{s_{\overline{p}p}}=200$\,GeV ~\cite{ua1ped}
where the same fit function, applied to the same $E_T^{jet}$ range,
results in the power $n=5.8$. 
Within the errors no
differences between $\gamma p$ and $\overline{p}p$
are found in this kinematic range.

In Fig.~\ref{sigmajet} the calculations of the PYTHIA generator
without multiple interactions (dotted curves) are compared with
the measured jet cross sections.
This model, and also HERWIG 
(not shown), show clear deficiencies in describing the cross
sections of jet production at large rapidities
and small transverse energies of the jets.
This kinematic region corresponds to events where the momentum fraction
of the parton from the photon is small $x_\gamma\approx 0.1$.

The PYTHIA calculation with multiple interactions provides a
fair description of the shape of the measured pseudo-rapidity cross section,
but gives a too large cross section in the region of small
$E_T^{jet} > 7$\, GeV (dashed curves).
The full curves represent the calculation of the PHOJET generator
which gives a good description of both the shape and the measured rates.
The results of the calculations including additional event energy
-- PHOJET and PYTHIA -- differ in absolute numbers in the small $E_T^{jet}$
region despite the use of the same QCD matrix elements and
structure functions.
However, the calculations include different modeling of
1) the beam remnant interactions ((semi-~) hard interactions in PYTHIA
   versus soft and (semi-~) hard interactions in PHOJET),
2) the transition to the non-perturbative
   soft scattering region (unitarization concept), and
3) the parton radiation effects (hard initial state parton radiation in
   PYTHIA, but not in PHOJET).
The comparison of these two models suggests that
inferences on the
parton content in the photon from the measured jet cross
sections have an uncertainty of about a factor of $2$
in the low $x_\gamma$ region.

The dash-dotted line in Fig.~\ref{sigmajet}b
indicates a PYTHIA calculation without
multiple interactions using a different
parameterization of the photon structure (LAC1 ~\cite{lac})
which has larger gluon content at small $x_\gamma$
than the parameterization used
for the other calculations in this paper (GRV-LO ~\cite{ggrv}).
This calculation also does not describe the measured energy density
(Fig.~\ref{etden}) and the jet cross section measurement
(Fig.~\ref{sigmajet}).
The same PYTHIA calculation with the LAC1 set including multiple
interactions gives,
after re-adjusting of the transverse momentum cut-off to
$p_t^{mia}\ge 2$\, GeV (compare section 5.1),
a jet cross section which is between $1.5<\eta^{jet}<2$
and above $E_T^{jet}>7$\, GeV a factor of two higher than the
calculation with the GRV-LO parton distribution function
in the photon, and factor 2.5 larger than the data.

The distribution of the energy density
outside the jets in the central rapidity region of the $\gamma p$
collision $-1\le\eta^*\le 1$, i.e. $1\le\eta\le 3$,
was used to correct the jet cross section as a function of pseudo-rapidity
above $\eta^{jet} > 1$ for effects of the underlying event:
the difference between the transverse energy density in the data and
the PYTHIA calculation without multiple interactions
(Fig.~\ref{etden}: dotted histogram)
indicates the event energy outside of jets which is produced
beyond the hard parton
scattering process, its parton showers and their fragmentation effects.
Using the information on the true parton momentum $x_\gamma$
in generated events
the pseudo-rapidity of the jet is correlated with $x_\gamma^{jets}$.
With the assumption that the calculations of the PYTHIA and HERWIG
generators represent a good approximation to the energy density
outside the jets arising from the hard parton scattering process,
the transverse jet energies are lowered by
the energy difference between the data and the calculation without
multiple interactions.
The corrections lower the cross section for jets between
$1 < \eta^{jet} <  2$ above
$E_T^{jet} >  7$\,GeV by $40\%$,
$E_T^{jet} > 11$\,GeV by $30\%$,
$E_T^{jet} > 15$\,GeV by $15\%$,
and demonstrate the strong influence of the
underlying event energy on the jet cross sections
(see also table ~\ref{incjetstab6}).

\begin{table}[b]
\caption[]
{\label{et_tab}
\small\it
  Measured differential $ep$ cross-section $d\sigma/dE_T$
  integrated over
  $-1.0 <  \eta <  2.5$ in the kinematical range defined by
  $Q^2<0.01~GeV^2$  and $0.3\le y \le 0.7.$}
 \vspace{5mm}
 \begin{center}
  \begin{tabular}{|l||c|c|c|} \hline
   $E_T~[GeV] $
&  $d\sigma/dE_T~[nb/GeV]$
&  $stat.\ error $
&  $syst.\ error $ \\
\hline \hline
        20--22   &  8.1   &      0.30   &     1.7    \\ \hline
        22--24   &  5.7   &      0.27   &     1.3    \\ \hline
        24--26   &  3.6   &      0.19   &     0.94   \\ \hline
        26--28   &  2.3   &      0.17   &     0.65   \\ \hline
        28--30   &  1.4   &      0.12   &     0.44   \\ \hline
        30--32   &  1.0   &      0.10   &     0.33   \\ \hline
        32--34   &  0.80  &      0.10   &     0.28   \\ \hline
        34--36   &  0.55  &      0.096  &     0.21   \\ \hline
        36--38   &  0.37  &      0.065  &     0.16   \\ \hline
        38--41   &  0.26  &      0.038  &     0.11   \\ \hline
        41--44   &  0.068 &      0.016  &     0.032  \\ \hline
        44--50   &  0.053 &      0.020  &     0.028  \\ \hline
   \end{tabular}
  \end{center}
 \end{table}

\begin{table}[tbp]
\caption[]
{\label{incjetstab1}
\small\it
  Measured differential $ep$ cross-section $d\sigma/dE_T^{jet}$
  for inclusive jet production integrated over
  $-1.0\leq \eta^{jet}\leq 2.0$ in the kinematical range defined by
  $Q^2<0.01~GeV^2$  and $0.25<y<0.7.$}
 \vspace{5mm}
 \begin{center}
  \begin{tabular}{|l||c|c|c|c|} \hline
   $E_T^{jet}~[GeV] $ &  $d\sigma/dE_T^{jet}~[nb/GeV]$ &
   $stat. error $&$syst. error $&$ syst.overall~error $ \\
\hline \hline
   ~7.0--~~9.0  & 20.8 &  0.56 & 4.4 &  5.4  \\
   \hline
   ~9.0-- 11.0  & 5.5  &  0.26 & 1.2 &  1.4  \\
   \hline
   11.0-- 13.0  & 1.74  &  0.12 & 0.37  &  0.45  \\
   \hline
   13.0-- 15.0  & 0.803 &  0.079 & 0.17  &  0.21 \\
   \hline
   15.0-- 17.0  & 0.283 &  0.044 & 0.059 &  0.074  \\
   \hline
   17.0-- 19.0  & 0.153 &  0.036 & 0.032  &   0.040  \\
   \hline
   19.0-- 21.0  & 0.0715 & 0.021 & 0.015  &  0.019   \\
   \hline
   21.0-- 25.0  & 0.0223 &  0.0081 & 0.0047 &  0.0058 \\
   \hline
   25.0-- 29.0  & 0.0172 &  0.0063 & 0.0036 &  0.0045 \\
   \hline
   \end{tabular}
  \end{center}

    \caption[]
{\label{incjetstab2}
\small\it
  Measured differential $ep$ cross-section $d\sigma/dE_T^{jet}$
  for inclusive jet production integrated over
  $-1.0\leq \eta^{jet}\leq 1.0$ in the kinematical range defined by
  $Q^2<0.01~GeV^2$  and $0.25<y<0.7.$}
 \vspace{5mm}
 \begin{center}
  \begin{tabular}{|l||c|c|c|c|} \hline
   $E_T^{jet}~[GeV] $ &  $d\sigma/dE_T^{jet}~[nb/GeV]$ &
   $ stat. error $&$ syst. error $&$  syst.overall~error $ \\
\hline \hline
   ~7.0--~~9.0  & 10.3 &  0.39 &  2.2  &  2.7
 \\
   \hline
   ~9.0-- 11.0  & 2.76 &  0.19 & 0.58  &  0.72
 \\
   \hline
   11.0-- 13.0  & 0.79 &  0.08 & 0.17 &  0.20
 \\
   \hline
   13.0-- 15.0  & 0.322 &  0.049 & 0.067 &  0.084
 \\
   \hline
   15.0-- 17.0  & 0.141 &  0.030 & 0.030  &  0.037   \\
   \hline
   17.0-- 19.0  & 0.0672 & 0.021 &  0.014 &  0.017   \\
   \hline
   19.0-- 21.0  & 0.0263 &  0.012 & 0.0055 &  0.0068   \\
   \hline
   \end{tabular}
  \end{center}

    \caption[]
{\label{incjetstab3}
\small\it
  Measured differential $ep$ cross-section $d\sigma/d\eta^{jet}$
  for inclusive jet production integrated over
  $E_T^{jet}>7.0~GeV$ in the kinematical range defined by
  $Q^2<0.01~GeV^2$  and $0.25<y<0.7.$}
 \vspace{5mm}
 \begin{center}
  \begin{tabular}{|l||c|c|c|c|} \hline
   $~~~~\eta^{jet}$   &  $d\sigma/d\eta^{jet}~[nb]$ &
   $ stat. error $&$ syst. error $&$  syst.overall~error $ \\
\hline \hline
   -1.0~--~~-0.5 &  5.9  & 0.64 & 1.3 & 1.5  \\
\hline
   -0.5~--~~~0.0 &  13.2 & 0.86 & 2.8 & 3.4   \\
\hline
    ~0.0~--~~~0.5 &  17.8 & 1.1 & 3.7 & 4.6   \\
\hline
    ~0.5~--~~~1.0 &  21.6 & 1.1 & 4.5 & 5.6   \\
\hline
    ~1.0~--~~~1.5 &  24.8 & 1.2 & 5.2 & 6.5   \\
\hline
    ~1.5~--~~~2.0 &  30.1 & 1.3 & 6.3 & 7.8   \\
\hline
  \end{tabular}
  \end{center}
 \end{table}

\begin{table}[tbp]
    \caption[]
{\label{incjetstab4}
\small\it
  Measured differential $ep$ cross-section $d\sigma/d\eta^{jet}$
  for inclusive jet production integrated over
  $E_T^{jet}>11.0~GeV$ in the kinematical range defined by
  $Q^2<0.01~GeV^2$  and $0.25<y<0.7.$}
 \vspace{5mm}
 \begin{center}
  \begin{tabular}{|l||c|c|c|c|} \hline
   $~~~~\eta^{jet}$   &  $d\sigma/d\eta^{jet}~[nb]$ &
   $ stat. error $&$ syst. error $&$   syst.overall~error $ \\
\hline \hline
    -0.5~--~~~0.0 &  1.23&  0.20& 0.26&  0.32  \\
\hline
    ~0.0~--~~~0.5 &  1.85&  0.24& 0.39&  0.48  \\
\hline
    ~0.5~--~~~1.0 &  2.36&  0.30& 0.50&  0.61  \\
\hline
    ~1.0~--~~~1.5 &  3.30&  0.35& 0.69 &  0.86  \\
\hline
    ~1.5~--~~~2.0 &  3.27&  0.33& 0.69 &  0.85  \\
\hline
  \end{tabular}
  \end{center}

    \caption[]
{\label{incjetstab5}
\small\it
  Measured differential $ep$ cross-section $d\sigma/d\eta^{jet}$
  for inclusive jet production integrated over
  $E_T^{jet}>15.0~GeV$ in the kinematical range defined by
  $Q^2<0.01~GeV^2$  and $0.25<y<0.7.$}
 \vspace{5mm}
 \begin{center}
  \begin{tabular}{|l||c|c|c|c|} \hline
   $~~~~\eta^{jet}$   &  $d\sigma/d\eta^{jet}~[nb]$ &
   $ stat. error $&$ syst. error $&$  syst.overall~error $ \\
\hline \hline
    ~0.0~--~~~0.5 & 0.50& 0.12 & 0.11 &  0.13  \\
\hline
    ~0.5~--~~~1.0 & 0.48& 0.12 & 0.10 &  0.13   \\
\hline
    ~1.0~--~~~1.5 & 0.67& 0.15 & 0.14 &  0.17   \\
\hline
    ~1.5~--~~~2.0 & 0.65& 0.14 & 0.14 &  0.17   \\
\hline
  \end{tabular}
  \end{center}

    \caption[]
{\label{incjetstab6}
\small\it
  The measured inclusive differential jet $ep$ cross-section
  $d\sigma/d\eta^{jet}$ after a subtraction of that part of
  the underlying event energy which is found above the
  calculated energy of the PYTHIA QCD generator without
  interactions of the beam remnants.
  The error bars reflect the statistical and systematic errors,
  added in quadrature.
  The kinematical range is defined by
  $Q^2<0.01~GeV^2$  and $0.25 < y < 0.7.$
}
 \vspace{5mm}
 \begin{center}
  \begin{tabular}{|c||c|c|} \hline
 $E_t^{jet}$ threshold & $d\sigma/d\eta^{jet}~[nb]$
                                  & $d\sigma/d\eta^{jet}~[nb]$ \\
 $[$GeV$]$             & for $1<\eta^{jet}<1.5$
                                  & for $1.5<\eta^{jet}<2$   \\ \hline\hline
  $7$                  & $14.1\pm 6.9$      & $17.8\pm 8.7$  \\ \hline
 $11$                  & $ 2.2\pm 1.0$      & $ 2.2\pm 0.9$  \\ \hline
 $15$                  & $0.54\pm 0.25$     & $0.52\pm 0.24$ \\ \hline
  \end{tabular}
  \end{center}
 \end{table}

\vspace*{1cm}
\noindent
{\bf Summary: Properties of jets and jet cross sections}

\noindent
The width of jets in photoproduction events is observed to decrease
with increasing jet energy.
Comparisons of jet width, measured in $\gamma p$ and $\overline{p}p$
scattering, show that these jets are of a common nature.

Jet cross sections are measured
as a function of the jet transverse energy
and pseudo-rapidity.
The shapes of the latter measurements are better described by LO QCD
calculations which include, in addition to the matrix elements of the
parton scattering processes and parton showers, interactions of
the beam remnants, than by calculations without such additional
interactions.
The strong influence of the underlying event energy on the measured
jet cross sections is demonstrated. This gives rise to
uncertainties in conclusions concerning
the parton scattering processes.

\section {Conclusions}

\noindent
In this paper detailed studies of high $E_T$
photoproduction events with center-of-mass energies around
$\sqrt{s_{\gamma p}}=200$\, GeV were presented.
The selected events have large transverse energy
$E_T > 20$\, GeV in the hadronic final state or exhibit
the production of jets with transverse energy above
$E_T^{jet} > 7$\, GeV summed in a cone of size $R=1$.
\begin{description}
\item{1.)}
    The differential cross section as a function of transverse energy
    was measured between $20\le E_T\le 50$\, GeV
    in the pseudo-rapidity range $-2.5\le\eta^*\le 1$
    (Fig.~\ref{etflow}a, Tab.~\ref{et_tab}).
    The shape is equally well described
    by a power law $E_T^{-n}$ with $n=5.9\pm 0.1$, and by an
    exponential function $e^{-\lambda E_T}$ with $\lambda=0.21\pm 0.01$.
\item{2.)}
    The transverse energy flow as a function of pseudo-rapidity is
    distinct from being flat and increases towards the origin of the
    $\gamma p$ center-of-mass system (Fig.~\ref{etflow}b).
\item{3.)}
    Both the multi-jet rate and the transverse energy contained in
    jets increase with the total transverse energy (Fig.~\ref{jetrate}).
\end{description}
QCD calculations which include, in addition to the hard parton
scattering, interactions between the beam remnants, provide
better descriptions of these transverse energy distributions than
models without such additional interactions.
\begin{description}
\item{4.)}
    The transverse energy density in the central region of the
    $\gamma p$ collision outside jets in resolved
    photon interactions reaches $1$\, ~GeV/rad per unit area in
    pseudo-rapidity and $\varphi$ space and is far higher than
    the energy density
    observed in direct photon processes, $0.3$\, GeV/rad
    (Fig.~\ref{etden}).
\item{5.)}
    The transverse energy depositions show short range correlations
    and long range anti-correlations relative to the transverse energy
    measured at the origin of the $\gamma p$ center-of-mass system
    (Fig.~\ref{correl}).
\end{description}
QCD models without interactions of the beam remnants provide too small
underlying event energy and show too strong energy correlations.
In QCD models with beam remnant interactions the additional energy
from the secondary interaction is essentially uncorrelated with
the primary hard parton interaction.
Such models not only provide the large underlying transverse
energy density observed in the data, but also reproduce the
energy correlations measured in the data.
This supports the models including interactions between the beam remnants,
although it is not inconceivable that a more complete QCD calculation
of the primary scattering process may reproduce the same effects.
\begin{description}
\item{6.)}
     The jets observed in $\gamma p$ collisions become denser
     around the jet axis with increasing jet energy (Fig.~\ref{jetprof}b).
\item{7.)}
     The jet cross section as a function of jet transverse energy
     decreases as $(E_T^{jet})^{-n}$ with $n=6.1\pm 0.5$
     between $7 < E_T^{jet} < 29$\, GeV
     in the pseudo-rapidity range $-1 < \eta^{jet} < 2$
     (Fig.~\ref{sigmajet}a, Tab.~\ref{incjetstab1},~\ref{incjetstab2}).
\item{8.)}
     The jet cross section was measured as a function of pseudo-rapidity
     (Fig.~\ref{sigmajet}b,
     Tab.~\ref{incjetstab3},~\ref{incjetstab4},~\ref{incjetstab5}).
     The strong influence of the treatment of the underlying event
     energy on the cross section was demonstrated
     (Tab.~\ref{incjetstab6}).
\end{description}
In order to draw conclusions on the parton scattering process,
QCD jet calculations have to be compared not only to the jet cross
sections (items 7,8), but also at least to the energy density of the
underlying event (item 4).

Comparisons of the $\gamma p$ results 1,2,4,6,7 with distributions from
$\overline{p}p$ collisions show that not only are the jets of
similar shape, but also the features of resolved $\gamma p$
interactions are similar to those of $\overline{p}p$ scattering.
Furthermore, the analysis of direct and resolved photon interactions
enabled, for the first time, the study of the underlying event energy
depending on the energy available to the photon remnant, revealing
new information on the multi-parton dynamics.

\vspace{0.5cm}
\noindent
{\bf Acknowledgments}

\noindent
We are indebted to J.Chyla, R.Engel and T.Sj\"ostrand
for help concerning the Monte Carlo event generators used in this analysis.
We are grateful to the HERA machine group whose
outstanding efforts made this experiment possible. We appreciate the
immense effort of the engineers and technicians who constructed and
maintain the detector. We thank the funding agencies for
their financial support of the experiment. We wish to thank the DESY
directorate for the hospitality extended to the non-DESY members
of the collaboration.

\end{document}